\documentclass[global,twocolumn,referee]{svjour}
\usepackage{graphicx}
\usepackage{siunitx}
\usepackage{tabularx}
\usepackage{booktabs}
\usepackage{hyperref}
\journalname{myjournal}
\begin{document}
\title{Fiber-laser driven gas-plasma based generation of THz radiation with 50 mW average power}
\author{Joachim Buldt\inst{1} \and Michael Mueller\inst{1} \and Henning Stark\inst{1} \and Cesar Jauregui\inst{1} \and Jens Limpert\inst{1,2,3}% etc
}                     % Do not remove
\institute{Institute of Applied Physics, Abbe Center of Photonics, Friedrich-Schiller-University Jena, Albert-Einstein-Str. 6, 07745 Jena, Germany \and Helmholtz-Institute Jena, Fröbelstieg 3, 07743 Jena \and Fraunhofer Institute for Applied Optics and Precision Engineering, Albert-Einstein-Str. 7, 07745 Jena, Germany}
\date{Received: 9 August 2019 / Accepted: 21 November 2019 / Published online: 26 November 2019}
\maketitle
\begin{abstract}
		We present on THz generation in the two-color gas plasma scheme driven by a high-power, ultrafast fiber laser system. The applied scheme is a promising approach for scaling the THz average power but it has been limited so far by the driving lasers to repetition rates up to 1 kHz. Here we demonstrate recent results of THz generation operating at a two orders of magnitude higher repetition rate. This results in a unprecedented THz average power of 50 mW. The development of compact, table-top THz sources with high repetition rate and high field strength is crucial for studying nonlinear responses of materials, particle acceleration or faster data acquisition in imaging and spectroscopy.
\end{abstract}
\section{Introduction}
\label{intro}
	The THz spectral region has been attracting growing interest over recent years thanks to a constantly increasing number of applications in many different fields. Examples for industrial applications are: monitoring compounding processes \cite{Jansen2010}, quality control by structural and chemical analysis \cite{Tonouchi2007}, quality inspection of food products as well as inspection of plastic weld joints \cite{Jansen2010}. In homeland security THz radiation is beneficial due to the transparency of many materials which enable the identification of hidden weapons by imaging \cite{Appleby2007} and explosives by fingerprinting \cite{Tonouchi2007,Ho2008}. For medical diagnostics the low photon-energy of THz radiations allows for ionization free morphological and compositional studies \cite{Ho2008} as well as for in vitro imaging of multiple types of tissue as well as in-vivo imaging of epidermal tissue \cite{Yang2016}. Scientific applications are particle acceleration \cite{Wong2013,Nanni2015,Kealhofer2016,Huang2016,Fallahi2016,Curry2018} and the study of ultrafast material dynamics \cite{Huber2012,Zaks2012,Schubert2014,Langer2018}. Of course, this is not an exhaustive list and the development of more powerful, broadband sources with high peak field strength enables even more applications and technologies in the THz spectral region.

One way to generate THz radiation is to use near-infrared lasers and transform the radiation into the THz region. 
Thus, the most straight-forward approach for average power scaling of the THz radiation would be to increase the average power of the driving laser source. However, individual limitations regarding the applicable laser parameters are given by the different frequency transformation schemes themselves.
One of these approaches is inspired by electronics and employs photo-conductive antennas (PCA). Hereby a short laser-pulse acts as an ultrafast switch that generates free carriers in a semiconductor. These free carriers are accelerated towards electrodes placed on the semiconductor, thus giving rise to a current that generates the THz radiation. Due to the materials employed in these antennas, the achievable THz power is limited by the damage threshold of the devices and until today the maximal reported average power is $ \SI{3.8}{\milli\watt} $ \cite{Yardimici2015}.

Approaches that rely on nonlinear conversion in crystals are optical rectification (OR) and difference frequency generation (DFG).
Both of these effects require phase-matching between the THz and the optical field in the nonlinear crystal, which can be accomplished by using different schemes \cite{Hebling2004,Blanchard2007,Vicario2014}. Using optical rectification in organic crystals \cite{Vicario2014} or lithium niobate \cite{Huang2013} allowed to achieve an infrared-to-THz conversion efficiency of more than $ \SI{3}{\percent} $, which is the highest value reached to date.
However, at a certain point the damage threshold of the crystal limits the average power scaling of the THz radiation and the to-date highest reported average power using this scheme is $ \SI{21.8}{\milli\watt} $ \cite{Huang2015}.

Approaches that rely on gases as the interaction medium are known as gas-plasma THz generation schemes. When first discovered \cite{Hamster1993,Hamster1994} the effect generating THz radiation was just ponderomotive and had a low efficiency ($ \ll 10^{-6} $). Later, several schemes to enhance the efficiency have been demonstrated \cite{Loeffler2000,Loeffler2002,Cook2000,Kress2006}. Out of those, the two-color plasma approach \cite{Cook2000} is one of the most popular ones nowadays due to its simple setup and a high achievable efficiency in the order of $ 10^{-4} $ as well as the large bandwidth and field strength. For this, a laser pulse is co-focused with its second harmonic, generating an asymmetry in the electric field which causes a much larger electron movement than the ponderomotive potential. However, the highest average power of THz radiation generated using this scheme so far was $ \SI{1.44}{\milli\watt} $ \cite{Oh2014}, with most experiments being limited by the average power of the driving laser system. Most of these experiments have been powered by Ti:sapphire lasers, which usually show a rather low average power due to the thermo-optical properties of the laser crystals.

In this contribution we present the latest results using the two-color plasma scheme driven by a state-of-the-art, tabletop, high-average-power, ultrafast ytterbium-fiber chirped pulse amplification system (Yb:FCPA) \cite{Muller16,Mueller2018}.

\section{Experimental Setup}
\begin{figure}[h]
	\centering
	\includegraphics[width=\columnwidth]{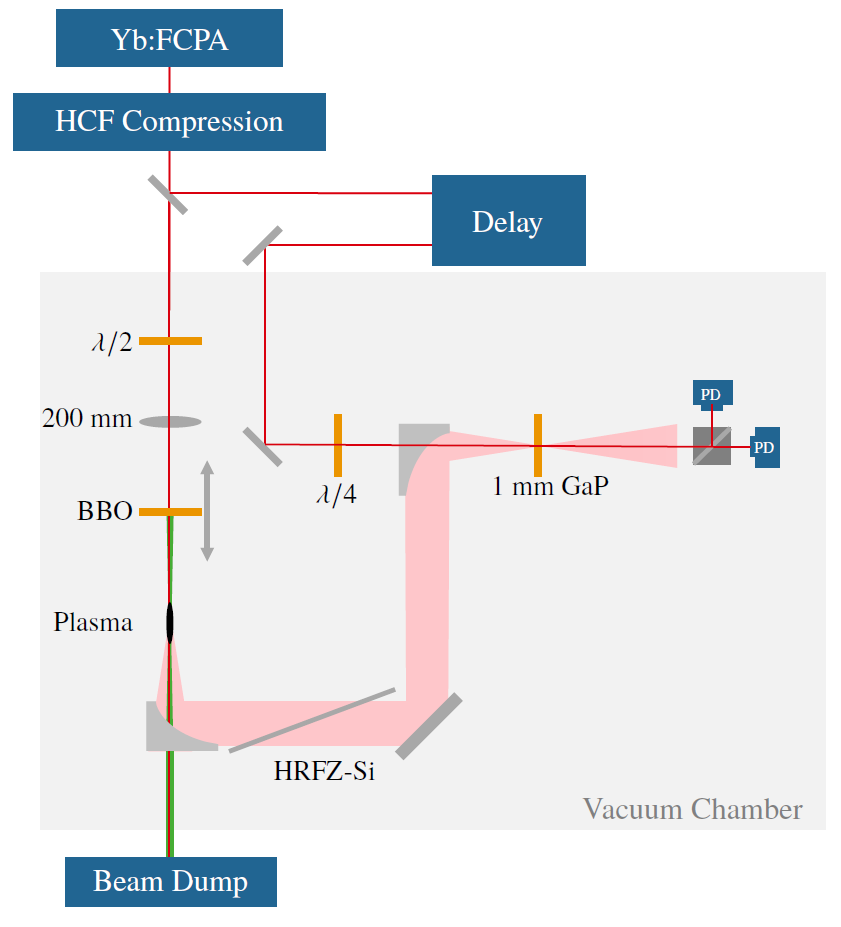}
	\caption[Setup]{Experimental setup for the THz generation: The pulses from the Yb:FCPA system are compressed with a HCF compression and focused with a $ \SI{200}{\milli\meter} $ lens through a $ \SI{100}{\micro\meter} $ thick BBO crystal into the gas. The laser light is dumped through a centered hole in the off-axis parabolic mirror that collimates the THz radiation. Residual light is blocked by a HRFZ-Si filter. For detection electro-optical sampling is done in GaP with a sample of the driving pulse.}
	\label{fig:setup}
\end{figure}
The experimental setup is depicted in Figure \ref{fig:setup}. The pulses were delivered by a 16-channel ytterbium-fiber chirped pulse amplification system (Yb:FCPA) \cite{Muller16,Mueller2018} at $ \SI{100}{\kilo\hertz} $ repetition rate with $ \SI{1.6}{\milli\joule} $ pulse energy and $ \SI{230}{\femto\second} $ pulse duration. Note that only two of the 16 available amplifier channels were used, so there is a great scaling potential when using the full available laser power. The pulses from the Yb:FCPA system are spectrally broadened in a hollow-core fiber (HCF) with $ \SI{350}{\micro\meter} $ core diameter and $ \SI{1}{\meter} $ length filled with $ \SI{1.5}{\bar} $ argon and compressed by chirped mirrors to $ \SI{30}{\femto\second} $ pulse duration afterwards. With a transmission of $ \SI{75}{\percent} $ through the hollow-core fiber, the remaining pulse energy was $ \SI{1.2}{\milli\joule} $. After compression the pulses enter a vacuum chamber filled with helium, nitrogen neon or argon at $ \SI{1}{\bar} $ absolute pressure. The beam is then focused with a lens of $ \SI{200}{\milli\meter} $ focal length through a $ \SI{100}{\micro\meter} $ thick BBO crystal to a $ \SI{40}{\micro\meter} $ focal spot (measured at low power without plasma). 

\begin{figure}[!h]
	\centering
	\includegraphics[width=\columnwidth,trim={0cm 0,0cm 0cm 0}]{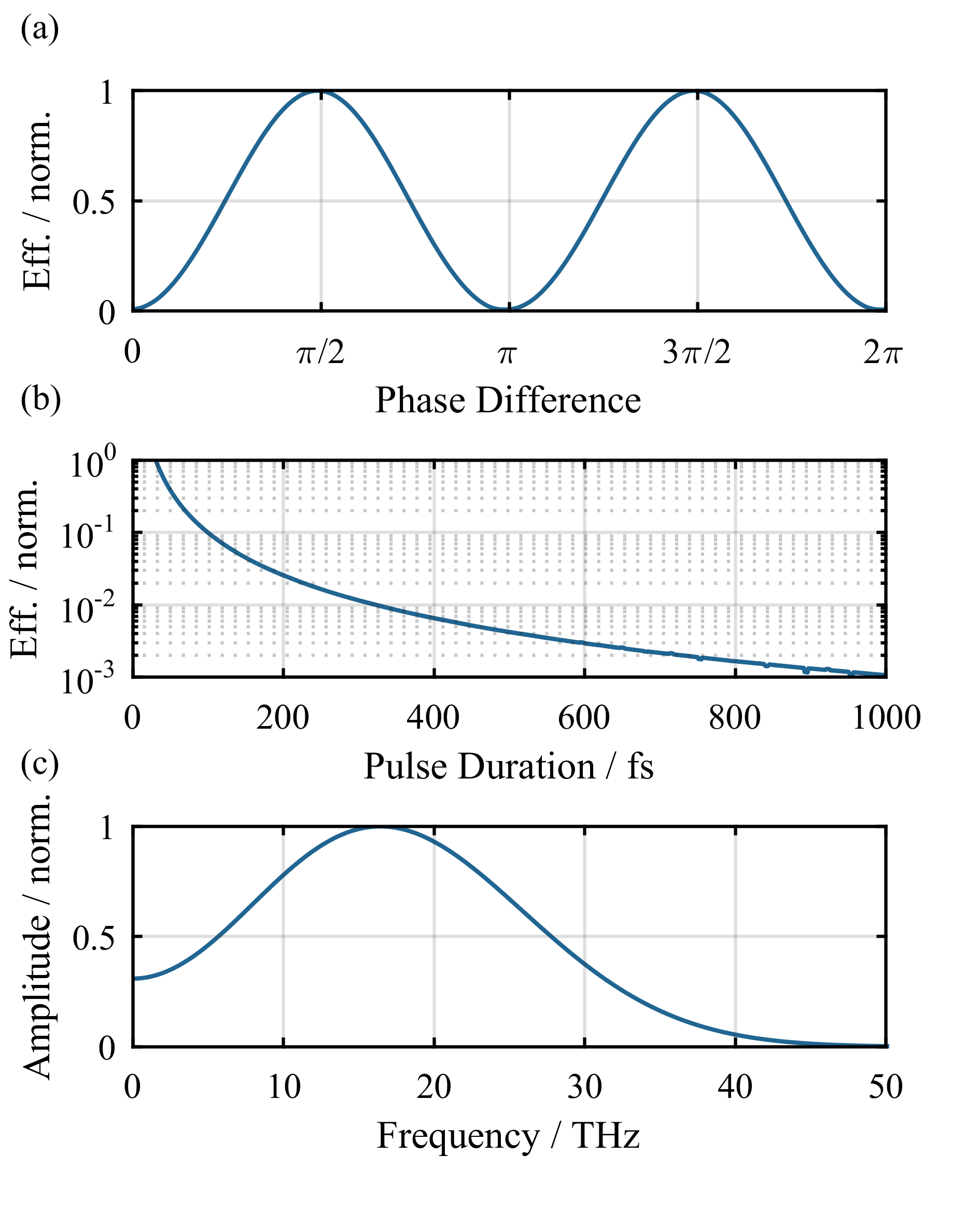}
	\caption{Simulation results based on the photo-current model \cite{Kim2009}. (a) Efficiency in dependence of the phase difference between the fundamental and second harmonic field. (b) Efficiency in dependence of the pulse duration, normalized to the efficiency of a 30 fs pulse. (c) THz spectrum simulated with the pulse parameters of the experiment and a phase difference of $ \pi/2 $ between the fundamental and second harmonic.}
	\label{fig:sim}
\end{figure}

From the simulation based on the photo-current model \cite{Kim2009} depicted in Figure \ref{fig:sim}a the phase difference between the fundamental and the second harmonic pulse is crucial for efficient THz generation in the two-color scheme. The propagation from the crystal to the focus through the gas allows to match the phase by adjusting the crystal position and taking advantage of the gas-dispersion.
The fundamental and its second harmonic are focused down and generate a plasma that gives rise to the THz radiation.
Afterwards, the THz beam is collected and collimated by a $ \SI{50.8}{\milli\meter} $ focal length, $ \SI{50.8}{\milli\meter} $ diameter, off-axis parabolic gold-mirror (OAP). The OAP has a hole in its center through which the remaining laser light is terminated onto a water-cooled beam dump outside of the vacuum chamber. The THz beam is additionally cleaned from residual laser radiation by  high-resistivity float-zone silicon (HRFZ-Si) filters with $ \SI{10}{\kilo\ohm\centi\meter} $ resistivity. 

For the detection of the THz radiation two options are available: 
In order to measure the average power of the THz radiation, an OAP can be inserted into the THz beam path to focus it through an optical chopper and onto a calibrated thermal power meter.
To minimize the influence of residual laser light on the power measurement three HRFZ-Si filters are inserted into the THz beam path and the transmission of each one of them is measured to calculate the generated THz power. 
To further ensure that the measured power really is the generated THz radiation each power measurement is validated by making a comparison measurement with long pulses. The chirp control of the Yb:FCPA system allows to quickly change the pulse duration to over $ \SI{1}{ps} $. From the simulation in \mbox{Figure \ref{fig:sim}b} one can see that the conversion efficiency drops by three orders of magnitude when going from $ \SI{30}{\femto\second} $ to $ \SI{1}{\pico\second} $ pulses. The average power reading with long pulses is subtracted from the one with short ones to eliminate any influence from residual infrared or stray light. 

As an addition to the power measurement the electrical field trace is measured by electro-optical sampling (EOS) \cite{Valdmanis1982-EOSampling}.
Electro-optical sampling is done by superimposing a small, delayed, circular polarized sample of the driving $ \SI{30}{\femto\second} $ pulse with the THz pulse in a $ \SI{1}{\milli\meter} $ thick gallium phosphate (GaP) crystal. Without the electric field of the THz radiation the difference of the signals of the two photo-diodes behind a polarizer vanishes. When the pulse delay is adjusted to overlap with the THz pulse, the electric field of the THz radiation causes a phase difference between the two polarization components of the probe pulse which is detected by the photo-diodes. By scanning the delay the temporal shape of the THz pulse can be retrieved. 
EOS is a commonly used way to characterize THz radiation  due to its simple setup, which is also the reason for its application in this experiment.
However it has to be noted, that it can not characterize the full spectrum of the radiation generated in this work, where the focus is showing the average-power and repetition rate scalability of the gas-plasma scheme.
%But due to the high repetition rate of the laser and a fast-moving delay-stage it allowed for a fast optimization of the experimental setup.

\section{Results}
\begin{figure}[!h]
	\centering
	\begin{minipage}{0.5\textwidth}
		\centering
		\includegraphics[width=\textwidth]{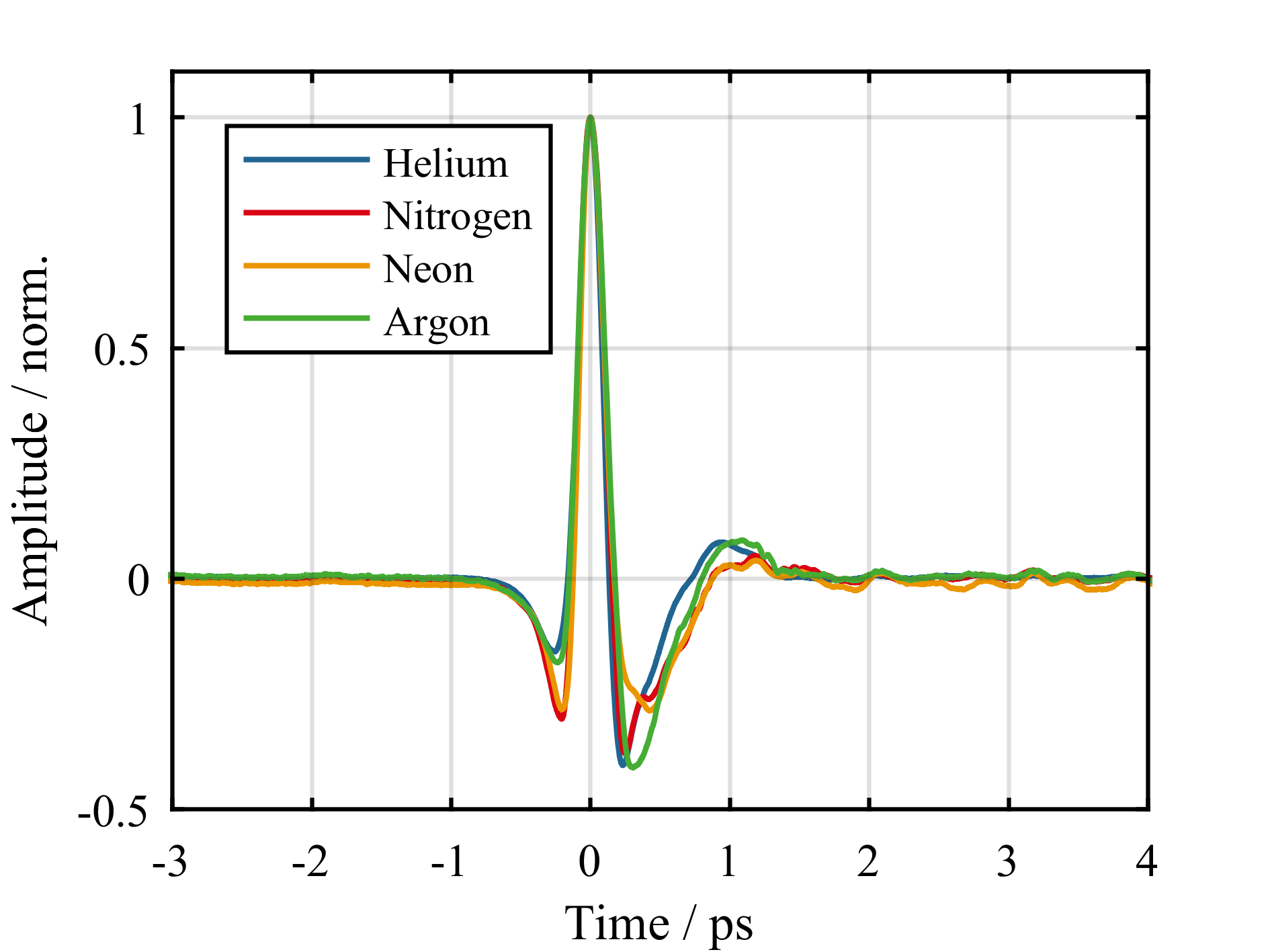} % first figure itself
		%\caption{Normalized electric fields measured by electro-optical sampling in GaP for four different gases.}
	\end{minipage}\hfill
	\begin{minipage}{0.5\textwidth}
		\centering
		\includegraphics[width=\textwidth]{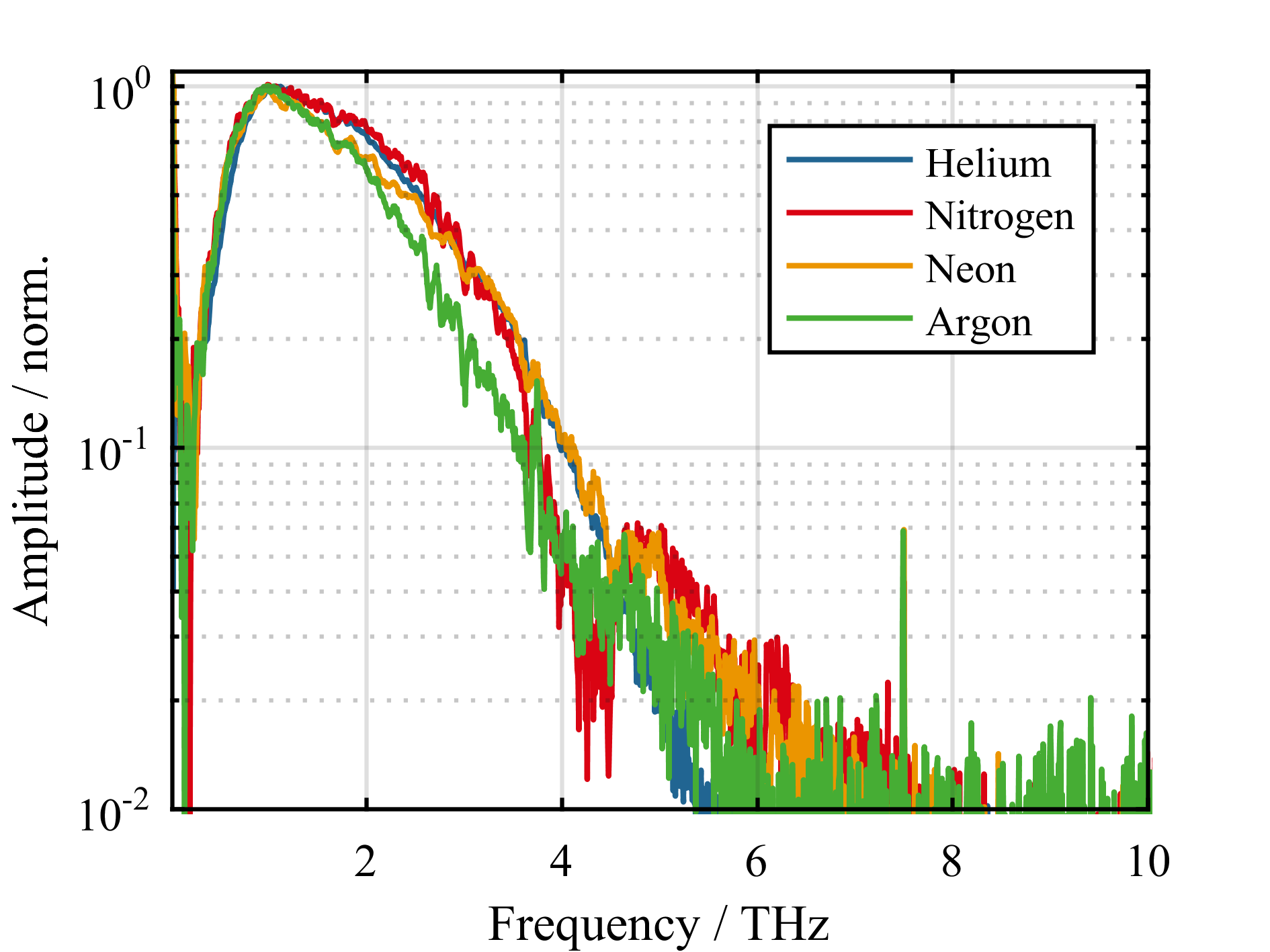} % second figure itself
		%\caption{Spectrum of the THz pulses retrieved by Fourier-transforming the EOS traces depicted in Figure \ref{fig:eos}. As can be seen in the simulation shown in the inset, the spectra are expected to extend to well over $ \SI{30}{\tera\hertz} $, being limited here by the bandwidth of the GaP-crystal used for the EOS measurements.}
	\end{minipage}
	\caption{Normalized electric fields measured by electro-optical sampling as well as the spectra of the THz pulses retrieved by Fourier-transforming the EOS traces. As can be seen in the simulation shown in Figure \ref{fig:sim}, the spectra are expected to extend to well over $ \SI{30}{\tera\hertz} $, being limited here by the bandwidth of the GaP-crystal used for the EOS measurements.}
	\label{fig:fft}
	\label{fig:eos}
\end{figure}
The results of the EOS measurements are depicted in \mbox{Figure \ref{fig:eos}}. The electrical field shows the single-cycle characteristic as it is expected from the applied generation scheme.
The measured spectra corresponding to the field-traces extend up to about $ \SI{5}{\tera\hertz} $, limited by the detection bandwidth of the EOS measurement. Due to the limited phase matching, decreasing nonlinear index towards higher frequencies as well as absorption of THz radiation in GaP \cite{Casalbuoni2008,Paradis2018}, the detection can not retrieve the full spectrum of the generated radiation. From the simulation depicted in \mbox{Figure \ref{fig:sim}c}, with the same pulse parameters as in the experiment a spectrum extending to well over $ \SI{30}{\tera\hertz} $
can be expected.
For the characterization of the full spectrum of the generated THz radiation, the implementation of a broadband detection system such as the air breakdown coherent detection (ABCD) scheme \cite{Dai2006} is planned for the future.
\begin{table}[!h]
	\flushleft 
	\caption[Power-measurements]{Average power of the THz radiation measured with the thermal powermeter as well as the calculated conversion efficiencies.}
	\label{tab:power}
	\centering
	\begin{tabularx}{\columnwidth}{cccc}
		\toprule
		\large\textbf{Gas} & & \large\textbf{Average Power}  & \large\textbf{Conv. Eff.} \\ \midrule			Helium             & & $ \SI{39}{\milli\watt} $ & $ 3.2\cdot 10^{-4} $      \\
		Nitrogen           & & $ \SI{47}{\milli\watt} $ & $ 3.8\cdot 10^{-4} $      \\
		Neon               & & $ \SI{50}{\milli\watt} $ & $ 4.2\cdot 10^{-4} $      \\
		Argon              & & $ \SI{39}{\milli\watt} $ & $ 3.2\cdot 10^{-4} $
		%			Helium             & & $ \SI{38.7}{\milli\watt} $ & $ 3.2\cdot 10^{-4} $      \\
		%			Nitrogen           & & $ \SI{46.7}{\milli\watt} $ & $ 3.8\cdot 10^{-4} $      \\
		%			Neon               & & $ \SI{49.8}{\milli\watt} $ & $ 4.2\cdot 10^{-4} $      \\
		%			Argon              & & $ \SI{38.7}{\milli\watt} $ & $ 3.2\cdot 10^{-4} $
	\end{tabularx} 
\end{table}

To evaluate the average THz power generated an additional measurement with a thermal powermeter was carried out for each gas. The measured values and the corresponding efficiencies of the THz generation process are shown in \mbox{Table \ref{tab:power}}. The highest power of $ \SI{50}{\milli\watt} $ was achieved using neon gas at an absolute pressure of $ \SI{1}{\bar} $ which corresponds to an conversion efficiency of $ 4.2 \cdot 10^{-4} $.

\section{Conclusion}
In conclusion, we have demonstrated the first step in the power-scaling of broadband, plasma-based THz sources by generating an average power of $ \SI{50}{\milli\watt} $. For the future a more precise temporal characterization of the THz radiation using an ABCD setup is planned as well as more systematic investigations on the experimental parameters which might further increase the efficiency.

For the results demonstrated here, only a fraction of the power delivered by our driving laser systems was required. By using the full power, similar pulse parameters could be achieved at over $ \SI{1}{\mega\hertz} $ repetition rate. With the same efficiency demonstrated here, an average THz power of $ \SI{0.5}{\watt} $ could be achieved in the near future.

\section*{Funding}
This project has received funding from the European Research Council (ERC) under the European Union's Horizon 2020 research and innovation programme (grant agreement No. 835306).

\section*{Acknowledgement}
J. Buldt acknowledges support from the IMPRS-PL within the international Ph.D. program.

\section*{Open Access}
This article is distributed under the terms of the Creative Commons Attribution 4.0 International License (\url{http://creativecommons.org/licen ses/by/4.0/}), which permits unrestricted use, distribution, and reproduction in any medium, provided you give appropriate credit to the original author(s) and the source, provide a link to the Creative Commons license, and indicate if changes were made.
% BibTeX users please use
\bibliographystyle{abbrv}
\bibliography{references}

\end{document}